# Effective Measurement Requirements for Network Security Management

Dr. Rabiah Ahmad
Department of System & Computer Communication
Universiti Teknikal Malaysia Melaka (UTeM), Melaka, Malaysia
Melaka, Malaysia
rabiah@utem.edu.my

Prof. Shahrin Sahib
Department of System & Computer Communication
Universiti Teknikal Malaysia Melaka (UTeM), Melaka, Malaysia
Melaka, Malaysia
shahrin@utem.edu.my

M.P. Azuwa
Research Department
Cybersecurity Malaysia
Selangor, Malaysia
azuwa@cybersecurity.my

*Abstract*— **Technical security metrics provide measurements in ensuring the effectiveness of technical security controls or technology devices/objects that are used in protecting the information systems. However, lack of understanding and method to develop the technical security metrics may lead to unachievable security control objectives and incompetence of the implementation. This paper proposes a model of technical security metric to measure the effectiveness of network security management. The measurement is based on the effectiveness of security performance for (1) network security controls such as firewall, Intrusion Detection Prevention System (IDPS), switch, wireless access point, wireless controllers and network architecture; and (2) network services such as Hypertext Transfer Protocol Secure (HTTPS) and virtual private network (VPN). We use the Goal-Question-Metric (GQM) paradigm [1] which links the measurement goals to measurement questions and produce the metrics that can easily be interpreted in compliance with the requirements. The outcome of this research method is the introduction of network security management metric as an attribute to the Technical Security Metric (TSM) model. Apparently, the proposed TSM model may provide guidance for organizations in complying with effective measurement requirements of ISO/IEC 27001 Information Security Management System (ISMS) standard. The proposed model will provide a comprehensive measurement and guidance to support the use of ISO/IEC 27004 ISMS Measurement template.**

## I. INTRODUCTION *(HEADING 1)*

Network security is defined as the security of devices, security of management activities related to the devices, applications/services, and end-users, in addition to security of the information being transferred across the communication links [2]. How much protection is required in ensuring the use of information and associated networks to conduct the business are well managed? How to identify and analyze network security controls to mitigate the network security risks? These questions have derived to implement and maintain secure and functional network is absolutely critical to the success of any organizationøs business operations [2][3]. Thus, it is important to measure network security effectiveness in handling the risks from the current threats, vulnerabilities and attacks.

According to [4], the practical challenges and issues are what to measure and what information to report in facilitates the senior management for any decision making. Obviously, the reported information is often based on what is easier to measure instead of what is actually meaningful strategically [5], [6], [7]. Does network security management is among the õeasierö information to measure?

Some organizations may be reported the measures from out of context perspective, without a baseline for comparison, or present simple measurements that do not show any kind of correlation, which greatly (or even completely) limits the value of the reported information [5][8].

### A. Requirements From ISO/IEC 27001 ISMS Standard

ISO/IEC 27001:2005 Information Security Management System (ISMS) [9] is intended to bring formal specification of information security under explicit management control. It is a mandated specific requirement, where organizations can therefore be formally audited and certified compliant with the standard.

The standard provides some confidence level of information protection among business organizations. With the existence of ISO/IEC 27001 ISMS certification, these organizations can increase their protection of information by having independent assessment conducted by the accredited certification body. The certificate has proven the potential marketing to the most business organizations, where a total of 7536 organizations have already been certified worldwide [10]. Obviously, there are other 27000 series that support this standard, including ISO/IEC 27002 Code of practice for information security management [11], ISO/IEC 27003 ISMS implementation guidance [12], ISO/IEC 27004 Information security management ó Measurement [13] and ISO/IEC 27005 Information security risk management [14].

There are 133 security controls in Annex-A of ISO/IEC 27001 ISMS standard. ISO/IEC 27002 [11] provides the best practice guidance in initiating, implementing or maintaining the security control in the ISMS. This standard regards that õnot all of the controls and guidance in this code of practice may be applicable and additional controls and guidelines not included in this standard may be required."

Information security measurement is a mandatory requirement in this standard where a few clauses are stated in [9]:

- "4.2.2(d) Define how to measure the effectiveness of the selected controls or groups of controls and specify how these measurements are to be used to assess control effectiveness to produce comparable and reproducible results;
- 4.2.3(c) Measure the effectiveness of controls to verify that security requirements have been met;
- 4.3.1(g) documented procedures needed by the organization to ensure the effective planning, operation and control of its information security processes and describe how to measure the effectiveness of controls;
- 7.2(f) results from effectiveness measurements; and
- 7.3(e) Improvement to how the effectiveness of controls is being measured."

Moreover, the new revision of ISO/IEC 27001:2013 [15] standard has also highlighted the importance of effective measurement in their mandatory requirement clauses 9 - Performance evaluation.

*B. Summary*

The standard highlighted that the organization must evaluate the information security performance and the effectiveness of the ISMS. The evaluation of the effectiveness should include but not limited to: (i) monitor and measure information security processes and controls; (ii) methods to use when monitor and analyze measurement for valid or significant result; (iii) time and personnel to perform the monitoring a nd measurement; (iv) determine time, duration and personnel to analyze the measurement results.

Thus, in ensuring the ISMS effectiveness, the information security measure can facilitate the management to make decision by the collection, analysis, evaluation and reporting of relevant performance-related measurements.

The importance of information security measurement is well defined and highlighted in both standards. Most of the research papers focused on information security metrics for general IT systems. However, lack of research on technical security metrics [16][17][18][19]. Thus, our research is focusing on the development of technical security measurement that will be incorporated in the technical security metric model.

## II. RELATED WORK

In understanding the requirements, the security metric, measure and effective measurement must be defined.

"Whatever the driver for implementing ISO 27001, it should no longer be just about identifying the controls to be implemented (based on the risk), but also about how each control will be measured. After all, if you can't measure it, how do you know it's working effectively?" [20].

In our previous study [21], we defined information security metrics is a measurement standard for information security controls that can be quantified and reviewed to meet the security objectives. It facilitates the relevant actions for improvement, provide decision making and guide compliancy to security standards. Information security measurement is a process of measuring or assessing the effectiveness of information security controls that can be described by the relevant measurement methods to quantify the data and the measurement results are comparable and reproducible.

Apparently, we also mapped the definitions of security metric, security measure and effective measurement from the previous studies [5][6][20][22][23][24][25][16][26][17][27][28][29][30][18][31][32][33][19][34] (refer to Table 1).

From Table I, we grouped the eight (8) components of security metrics and supported by the components in security measures. The definitions of security metric and security measures are quite similar through the analysis of the descriptions. To ease the understanding, the metric is also sometimes called a "measure" [27]. However, in the development of TSMM, we intend to develop a security metric that can consist of a few security measures.

We also derived the eight (8) criteria of the effective security metric (ESM) that are supported by the following statement:

a) Meet security objectives - ESM should gauge how well organization is meeting its security objectives. It should also have a clearly defined set of variables which are acceptable, unacceptable and excellent range of values that can be easily identified by the audience to which the measure is communicated.

b) Quantifiable values – ESM should be a quantitatively measurable that derived from precise and reliable numeric values and expressed by using understood and unambiguous units of measure.

c) Simple measurement – ESM should be easily recognize and comprehended by the audience for which they are intended. The measurement method should be produced by a process or procedure to collect data, determine the data source, scale or score, analysis, and reporting of relevant data. The right and competent personnel should be identified to conduct the measurement and able to analyze and produce the accurate report.

TABLE I.  A MATRIX MAPPINGS THE DEFINITIONS OF SECURITY METRICS, SECURITY MEASURES AND EFFECTIVE MEASUREMENT

| Security Metric | Security Measure | Effective Measurement |
|---|---|---|
| (1) Security Objectives<br>• Identify the adequacy of security controls | • Clearly defined acceptable value<br>• Performance goals and objectives (efficiency, effectiveness) | • Meet security objectives and requirements<br>• Clearly defined |
| (2) Quantifiable, computed value | • Quantifiable information<br>• Scope of measurement (Process, performance, outcomes, quality, trends, conformance to standards and probabilities) | • The value is objective and quantifiable<br>• Determine the Key-Performance-Indicator (KPI) |
| (3) Method of Measurement<br>• Process of data collection, data from security assessment process | • Easily identified<br>• Quantitative indications by some attributes of a control or process | • Simple measurement<br>• Low cost and easy access<br>• Capability to measure accurately |
| (4) Analysis of Data<br>• Comparable to a scale/benchmark/Predetermined baseline<br>• Repeatable | • Apply formulas for analysis<br>• Track changes<br>• Quantifiable information for comparison | • Consistent value<br>• Accurate time and data<br>• Comparable and reproducible results<br>• Security controls are implemented correctly, operating as intended, and meeting the desired outcome. |
| (5) Security Indicator/Characteristics<br>• Meaningful result (score, rating, rank, or assessment result) | • Monitor the accomplishment | • Increase confidence level<br>• Security improvement |
| (6) Reporting relevant data | • Communicated/Reported<br>• Intended audience | • Present to targeted audience/Stakeholder |
| (7) Decision making | • Facilitate decision making | • Facilitate corrective action |
| (8) Requirement to Standard, regulatory, financial and organizational reasons | | • Align with business goals and regulations |

d) Comparable result – ESM should produce a baseline for comparison purposes, repeatable or consistently reproducible, so that different people at different times can make the same measurement. Apparently, this supports the adequacy of in-place security controls, policies, and procedures; security controls are implemented correctly, operating as intended, and meeting the desired outcome.

e) Corrective action - ESM should provide the appropriate timeliness and frequency of measurement for the change of measurement target so that the latency of measures does not defeat their purpose. ESM should be collected and reported in a consistent manner. ESM should provide the management to decide the new investment in additional information security resources, identify and evaluate non-productive security controls, and prioritize security controls for continuous monitoring.

f) Targeted audience/Stakeholder – ESM should be easily identified by the audience/stakeholder to whom the measure is communicated. For example, provide the relevant measures that produce the significant result for the management to make decision.

g) Security Improvement – ESM should provide some indicators that could be a sign of relevant security characteristics that prescribes the meaning of obtained security values and achieves to some level of improvement.

h) Align with business goals - ESM should provide a benefit to the business it supports.

The development of our TSMM is based on the above criteria and to focus on security performance for the relevant controls (see Fig.1).

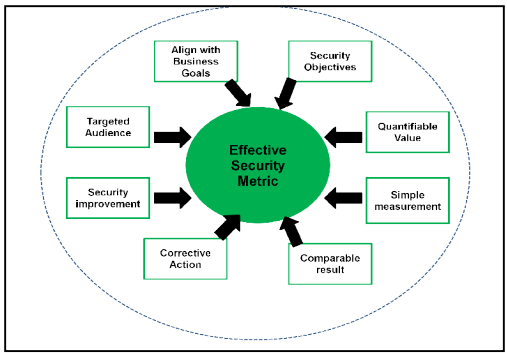

Figure 1. Eight Criteria of Effective Security Metric

### III. RESEARCH METHOD FOR DEVELOPMENT OF TECHNICAL SECURITY METRIC MODEL (TSMM)

The GQM approach was originally developed by Basili et.al [1] in evaluation and measurement of software products and development processes. Ever since developed, this approach was used consistently focus on the software measurement and processes [35]. There were also a few research studies on business processes [36][37][38] and security metrics [26][39][40][41][42][43][44]. However, there is no research study conducted for measuring the network security management using the same approach.

To achieve the objective of developing the TSMM, we propose a research method based on a combination of approaches. The outcome of this research method is the introduction of network security management metrics as attributes to the TSMM.

The first approach is to define the technical security metric (TSM). We set our goal to meet the requirements from ISO/IEC 27001 ISMS standard. The paradigm of Goal-Question-Metric (GQM) [1] is used and described further which to align with standard requirement (Fig.2).

We combine the developed Goal-Question-Metric (GQM) paradigm and data of literature review (Fig.3) as a first step. This approach is used for developing the initial TSM in a top-down manner, from general objective to the relevant metrics or outputs and combines the inputs from the literature review. The application results in GQM models, leading to the initial TSMM. However, this initial development work remains subjective and potentially incomplete.

In the second approach (Step 2), we use the GQM method consists of four phases [45]: planning, definition, data collection, interpretation (see Fig.4). The explanation of these phases is based on the compliancy to the requirement controls of ISO/IEC 27001 standard [9] for A.10.6 Network security management (NSM); A.10.6.1 Network controls; and A.10.6.2 Secure network services.

Our implementation adopts the processes and activities by [41] and [46].

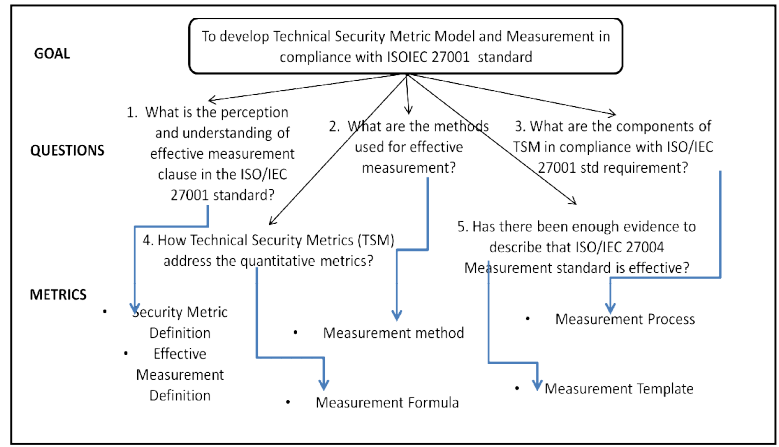

Figure 2. Eight Criteria of Effective Security Metric

- The *Planning phase*: The NSM-team is established and the compliance requirement is clearly delivered. The desired improvement areas such as performance, security and monitor are identified. The team selects and characterizes the products or controls to be studied. The result of this phase is a project plan that outlines the characterization of the products or controls, the schedule of measuring, the organizational structure, and necessary awareness and training for people involved in measurements.

- The *Definition phase*: The measurement goals are defined. This phase is also to identify and analyze the perception and understanding of effective measurement requirement from ISO/IEC 27001 standard [9]. We will create a new template to gather all related information based on some other templates from ISO/IEC 27004 [13] and NIST SP800-55 [27]. For the purpose of this, the interviews may be conducted with people (management and technical) involved in the process or product under study. Based on the goals, relevant questions are developed to identify the specific quality attributes and to re-define the goals precisely. For each question a hypothesis with an expected answer should be defined. Next, the metrics are defined for each question and checked on consistency and completeness. Results of this phase are an analysis of compliance plan and a measurement plan.

- The *Data Collection phase* – the team is required to prepare the data collection within their knowledge and availability. The data may be extracted manually or electronically and may involve automated data collection tools. Results of this phase are to develop the data support system consisting of spreadsheets, statistical tools, database applications and presentation tools.

- The *Interpretation phase* - the collected data is processed and analyzed according to the metrics defined. The measurements result should be able to answer the questions, and with the answers it can be evaluated if the initial goals are attained. Moreover, the measurement result should provide some values that describing the performance measurement of the security controls.

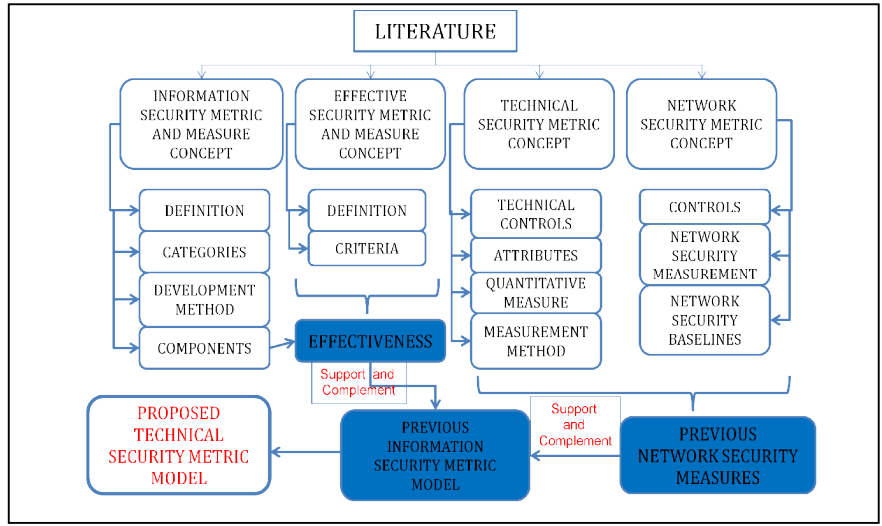

Figure 3. Data from literature review

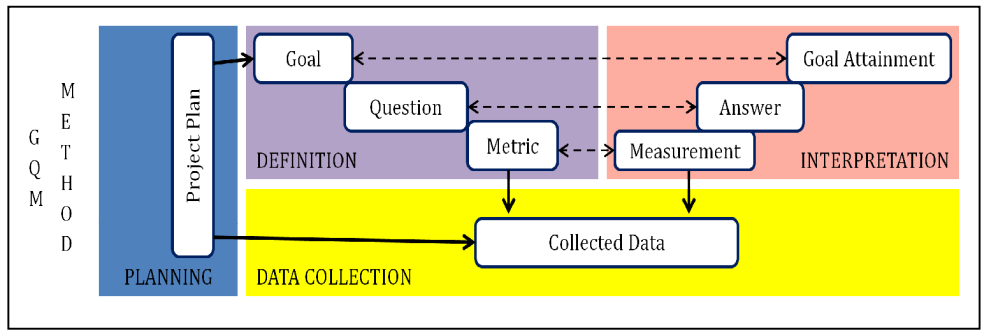

Figure 4. The four phases of GQM-method [45]

The second approach is used as a validation/improvement of the first step. It is based on a literature review of security metric standards and guidelines and measurement methods for network security controls. This approach is a bottom-up, being an analysis of the literature to identify the metrics currently used. A comparative analysis is developed between the metrics and those defined through GQM. This comparison is summarized in an analysis table.

As shown in Fig.5, we map the GQM-method with ISO/IEC 27004 template for an information security measurement construct and show the synchronization link (relevant colored-box). We refer to this standard as a reference and example to form a GQM-Measurement plan.

Once the literature is completely surveyed, the development of GQM-Measurement plan should be ready. The relevant people should be interviewed to validate the initial TSMM. Finally, the TSMM is accordingly revised.

*A. GQM-Measurement Plan*

We develop a GQM-Measurement plan consists of goals, questions, and metrics in a hierarchical structure (see Fig. 6) based on [1][45].

In developing the goals, the security objectives of A.10.6, A.10.6.1 and A.10.6.2 of ISO/IEC 27001 requirement controls [9] are referred. At this stage, the understanding of the security control requirements is very important. The understanding can be obtained through the interview with the relevant people and checking available process or product descriptions [46]. If goals are still unclear, a reference to ISO/IEC 27002 [11], FDIS ISO/IEC 27033 [2] and NIST SP800-55 [25] can also assist.

The proposed questions shall refine the goals make them operational enough so that it would not create difficulties to reveal the relationship to the collected data and ease the interpretation of the answers towards the goals [46]. The questions are also derived from the literature reviews.

The questions are stated in a quantitative way where data can be collected by measurements. We provide the expected answers to the questions and formulated as hypotheses. Through hypotheses, we can learn the effect from measurements and compare the knowledge before and after measurements.

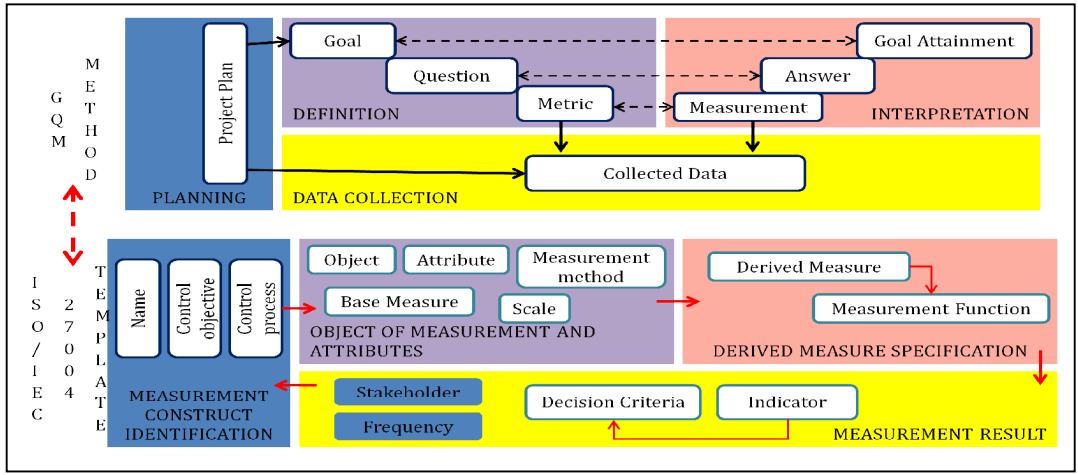

Figure 5. Synchronization between GQM-Method and ISO/IEC 27004 Measurement Template

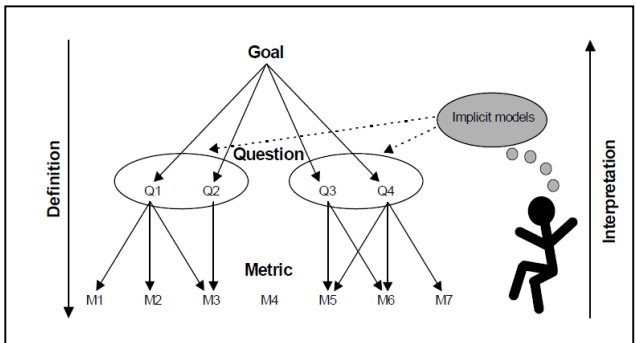

Figure 6. The GQM Paradigm by Basili et.al [1]

According to [1][41][46], we can define several metrics for each question. It is also possible that one metric may be used to answer different questions under the same goal. We choose metrics with quantitative level making it possible to assign numbers to a quality attribute. Metrics are defined to answer the relevant questions and should be able to support or reject the stated hypotheses (if any).

A simple Goal-Measurement plan is developed for the purpose of this discussion (as full development of plan is currently in progress). The example of GQM-Measurement plan as stated in Table II.

TABLE II. EXAMPLE OF GQM-MEASUREMENT PLAN

| Goal | G1 | A.10.6.1 Network controls - Networks shall be adequately managed and controlled, in order to be protected from threats, and to maintain security for the systems and applications using the network, including information in transit. |
|---|---|---|
| Question | Q1 | What are the risk levels for network controls and security controls that protect your information? |
| Metric | M1.1 | Risk Assessment = Asset Value x Threat x Vulnerability |
| Question | Q2 | What are the monitoring mechanisms that your organization has? |
| Metric | M2.1 | Frequency of audit logging review |
| Metric | M2.2 | Security Incidents report (IDS/IPS/user report) - Comparison of number of total incidents with the threshold. |
| Question | Q3 | How often the security assessment and/or penetration testing are conducted within a year? |
| Metric | M3.1 | Frequency of assessment conducted |
| Metric | M3.2 | Success or failure rate for corrective action |
| Metric | M3.3 | Conducted by trained/experience staff |
| Question | Q4 | How often your organization conduct the security updates for network controls? |
| Metric | M4.1 | Success and failure rates of security updates |
| Metric | M4.2 | Frequency/periodic of maintenance |
| Question | Q5 | Who is responsible to ensure the effectiveness of network controls is intact? |
| Metric | M5.1 | Rate of understanding the job description |
| Metric | M5.2 | Qualification, Training and Education attended |
| Question | Q6 | What are the authentication mechanism in accessing the network and systems used in your organization? |
| Metric | M6.1 | Password quality – manual (Number of passwords which satisfy organization's password quality policy for each user) |
| Metric | M6.2 | Password quality - automated |
| Metric | M6.3 | Number of password being shared? |
| Metric | M6.4 | Ratio of passwords crackable within 4 hours. |
| Question | Q7 | Who is responsible to ensure the effectiveness of network controls is intact? |
| Metric | M7.1 | Rate of understanding the job function |
| Metric | M7.2 | Qualification, Training and Education attended |
| Metric | M7.3 | Ratio of responsible personnel to total number of staff |
| Question | Q8 | What are the mechanism used to authorize the relevant users to access the networks and systems? |
| Metric | M8.1 | Number of restricted access methods (network segment, IP address, MAC address, firewall, etc.) |

## IV. CONCLUSION AND FUTURE WORK

The objective of this paper is to identify and to define a set of metrics for the TSMM with a systematic and scientific approach to comply with ISO/IEC 27001 standard. We use the GQM approach on the TSMM and review with regards to the literature. The result of this paper is the enrichment of the TSMM with suited network security management metrics.

Although the initial developed TSMM are validated through literature analysis, their testing in a real case would provide a concrete instantiation and validation of their relevance. The GQM-Measurement plan is currently being developed to suit the security objectives. The validation will be conducted with the network security experts.

As part of the next step of our future work, the metrics will be integrated into the initial TSMM and a case study is to be conducted using our GQM-Measurement plan. This will validate the final TSMM.

## ACKNOWLEDGMENT

The authors wish to acknowledge and thank members of the research teams of the Long Term Fundamental Research Grant Scheme (LRGS) number LRGS/TD/2011/UKM/ICT/02/03 for this work. The research scheme is supported by the Ministry of Higher Education (MOHE) under the Malaysian R&D National Funding Agency Programme. This project is also supported by the CyberSecurity Malaysia and the Universiti Teknikal Malaysia Melaka (UTeM), Malaysia.

## AUTHORS PROFILE

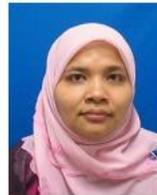

**Rabiah Ahmad** received Ph.D in Health Informatics at University of Sheffield (UK) and Master of Science in Information Security from Royal Holloway University of London (UK). She is currently appointed as Associate Professor at Universiti Teknikal Malaysia Melaka (UTeM) and acting as Deputy Director at Centre for Research Innovation and Management. Rabiah Ahmad involved with various research in information security and health informatics. She has become project leader for 4 research projects funded by the Ministry of Science, Technology and Innovation, Malaysia and the Ministry of Higher Education, Malaysia. She has wrote 3 academic books, more than 5 chapters in books, 30 articles in International Indexed Journal, 2 articles In Local Indexed Journal, more than 30 in International Proceedings and 3 in Local Proceedings. She has been invited as Manuscript Reviewer for several International Journals such as International Journal of Medical Informatics, International Journal on Cryptography and Journal of Soft Computing.

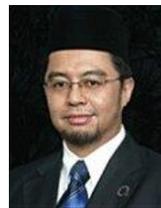

**Shahrin Sahib** received the Bachelor of Science in Engineering, Computer Systems and Master of Science in Engineering, System Software in Purdue University in 1989 and 1991 respectively. He received the Doctor of Philosophy, Parallel Processing from University of Sheffield in 1995. His research interests include network security, computer system security, network administration and network design. He is a member panel of Expert National ICT Security and Emergency Response Centre and also Member of Technical Working Group: Policy and Implementation Plan, National Open Source Policy. He is a Professor and Deputy Vice Chancellor Office (Academy and International) at Universiti Teknikal Malaysia Melaka (UTeM).

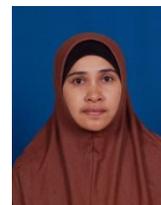

**M.P Azuwa** is the Specialist of CyberSecurity Malaysia, an agency under the Ministry of Science, Technology and Innovation, Malaysia. Azuwa holds a Masterøs degree in Computer Science from the Universiti Putra Malaysia, Malaysia and a Bachelorøs degree in Computer Science from the same university. She is the holder of Certified Information Security Manager (CISM) Certified in Risk and Information Systems Control (CRISC) from ISACA, USA; Professional on Critical Infrastructure Protection (PCIP) from Critical Infrastructure Institure (CII), Canada; Certified SCADA Security Architect (CSSA) from InfoSec Institute, USA; Certified BS7799 Lead Auditor ó Information Security Management System (ISO/IEC 27001); SANS GIAC Security Essential Certified (GSEC). Azuwa has been awarded Information Security Practitioner Honouree in July 2011 by the (ISC)2, USA. She has contributed various publications and presented papers on topics related to vulnerability assessment, SCADA security and information security management. She is currently pursuing his PhD at the Universiti Teknikal Malaysia Melaka (UTeM), Malaysia.